\documentclass{svmult}

\usepackage{grossdruck}      % switches to 11pt style - output same as
\usepackage{makeidx}         % allows index generation
\usepackage{graphicx}        % standard LaTeX graphics tool
\usepackage{multicol}        % used for the two-column index
\usepackage[bottom]{footmisc}% places footnotes at page bottom
\usepackage{graphics}
\usepackage{amsmath}
\usepackage{amsfonts}
\usepackage{amssymb}

\makeindex             % used for the subject index
\begin{document}

\title*{Updating Probabilities: An Econometric Example\thanks%
{Presented at the 3rd Econophysics Colloquium, Ancona, Italy, Sept 27-29, 2007.
}}
\author{Adom Giffin\inst{}}
\institute{Department of Physics, University at Albany-SUNY, Albany, NY 12222, USA
\texttt{ physics101@gmail.com}}
\maketitle

\begin{abstract}
We demonstrate how information in the form of observable data and moment
constraints are introduced into the method of Maximum relative Entropy (ME).
A general example of updating with data and moments is shown. A specific
econometric example is solved in detail which can then be used as a template
for real world problems. A numerical example is compared to a large
deviation solution which illustrates some of the advantages of the ME method.
\end{abstract}

\section{Introduction}

\label{intro} The MaxEnt method \cite{Jaynes57} was designed to assign probabilities. This
method has evolved to a more general method, the method of Maximum
(relative) Entropy (ME) \cite{ShoreJohnson80,Skilling88,CatichaGiffin06}
which has the advantage of not only assigning probabilities but \emph{%
updating} them when new information is given in the form of constraints on
the family of allowed posteriors. The main purpose of this paper is to show
both general and specific examples of how the ME method can be applied using
data and moment constraints.

The two preeminent updating methods are the ME method and Bayes' rule. The
choice between the two methods has traditionally been dictated by the nature
of the information being processed (either constraints or observed data) but
questions about their compatibility are regularly raised. Our first
objective is to review how data is introduced into the ME method.

Next we show a general example of updating with two different forms of
information: moments and data. The solution resembles Bayes' Rule. The
difference between this solution and the traditional Bayes form results from
using the moment constraint. This constraint modifies the usual Bayesian
likelihood. In an effort to put some names to these pieces we will call the
standard Bayesian likelihood the \emph{likelihood} and the part associated
with the moment the \emph{likelihood modifier} so that the product of the
two yields the \emph{modified likelihood}. We extend this general example by
solving a specific ill-behaved econometric problem in detail, which can then
be used as a template for real world problems. Numerical solutions are
produced to explicitly illustrate the case.

Recently, ill-behaved problems have been solved using large deviation
theory or information-theoretic approaches. All of these methods have a
common premise: they rely on asymptotic arguments. The ME method does not
need such assumptions to work and therefore can process finite amounts of
data well. However, when ME is taken to asymptotic limits one recovers the
same solutions that the information-theoretic methods produce. This is
discussed by comparing the numerical solution to our specific example and
the solution that is attained by the method of types \cite{CoverThomas}.

\section{Updating with data using the ME method}

\label{sec:2} Our first concern when using the ME method to update from a
prior to a posterior distribution is to define the space in which the search
for the posterior will be conducted. We wish to infer something about the
values of one or several quantities, $\theta \in \Theta $, on the basis of
three pieces of information: prior information about $\theta $ (the prior),
the known relationship between $x$ \emph{and} $\theta $ (the model), and the
observed values of the data $x\in \mathcal{X}$. Since we are concerned with
both $x$ \emph{and} $\theta $, the relevant space is neither $\mathcal{X}$
nor $\Theta $ but the product $\mathcal{X}\times \Theta $ and our attention
must be focused on the joint distribution $P(x,\theta )$. The selected joint
posterior $P_{\text{new}}(x,\theta )$ is that which maximizes the entropy,%
\begin{equation}
S[P,P_{\text{old}}]=-\int dxd\theta ~P(x,\theta )\log \frac{P(x,\theta )}{P_{%
\text{old}}(x,\theta )}~
\end{equation}%
subject to the appropriate constraints. $P_{\text{old}}(x,\theta )$ contains
our prior information which we call the \emph{joint prior}. To be explicit,%
\begin{equation}
P_{\text{old}}(x,\theta )=P_{\text{old}}(\theta )P_{\text{old}}(x|\theta )~,
\end{equation}%
where $P_{\text{old}}(\theta )$ is the traditional Bayesian prior and $P_{%
\text{old}}(x|\theta )$ is the likelihood. It is important to note that they 
\emph{both} contain prior information. The Bayesian prior is defined as
containing prior information. However, the likelihood is not traditionally
thought of in terms of prior information. Of course it is reasonable to see
it as such because the likelihood represents the model (the relationship
between $\theta $ and $x)$ that has already been established. Thus we
consider both pieces, the Bayesian prior and the likelihood to be \emph{prior%
} information.

The new information is the \emph{observed data}, $x^{\prime }$, which in the
ME framework must be expressed in the form of a constraint on the allowed
posteriors. The family of posteriors that reflects the fact that $x$ is now
known to be $x^{\prime }$ is such that%
\begin{equation}
P(x)=\int d\theta ~P(x,\theta )=\delta (x-x^{\prime })~.
\label{data constraint}
\end{equation}%
This amounts to an \emph{infinite} number of constraints: there is one
constraint on $P(x,\theta )$ for each value of the variable $x$ and each
constraint will require its own Lagrange multiplier $\lambda (x)$.
Furthermore, we impose the usual normalization constraint, 
\begin{equation}
\int dxd\theta ~P(x,\theta )=1~.
\end{equation}

Maximize $S$ subject to these constraints, 
\begin{equation}
\delta \left\{ 
\begin{array}{c}
S+\alpha \left[ \int dxd\theta P(x,\theta )-1\right] \\ 
+\int dx\lambda (x)\left[ \int d\theta P(x,\theta )-\delta (x-x%
%TCIMACRO{\U{b4}}%
%BeginExpansion
{\acute{}}%
%EndExpansion
)\right]%
\end{array}%
\right\} =0~,
\end{equation}%

and the selected posterior is 
\begin{equation}
P_{\text{new}}(x,\theta )=P_{\text{old}}(x,\theta )\,\frac{e^{\lambda (x)}}{Z%
}~,  \label{solution a}
\end{equation}%
where the normalization $Z$ is 
\begin{equation}
Z=\,e^{-\alpha +1}=\int dxd\theta \,P_{\text{old}}(x,\theta )\,e^{\lambda
(x)}~,  \label{Z}
\end{equation}%
and the multipliers $\lambda (x)$ are determined from (\ref{data constraint}%
), 
\begin{equation}
\int d\theta ~P_{\text{old}}(x,\theta )\frac{\,e^{\lambda (x)}}{Z}=P_{\text{%
old}}(x)\frac{\,e^{\lambda (x)}}{Z}=\delta (x-x^{\prime })~.
\end{equation}%
Therefore, substituting $e^{\lambda (x)}$ back into (\ref{solution a}), 
\begin{equation}
P_{\text{new}}(x,\theta )=\frac{P_{\text{old}}(x,\theta )\,\delta
(x-x^{\prime })}{P_{\text{old}}(x)}=\delta (x-x^{\prime })P_{\text{old}%
}(\theta |x)~.
\end{equation}

The new marginal distribution for $\theta $ is%
\begin{equation}
P_{\text{new}}(\theta )=\int dxP_{\text{new}}(x,\theta )=P_{\text{old}%
}(\theta |x^{\prime })~.
\end{equation}%
This is the familiar Bayes' conditionalization rule. To summarize: $P_{\text{%
old}}(x,\theta )=P_{\text{old}}(x)P_{\text{old}}(\theta |x)$ is updated to $%
P_{\text{new}}(x,\theta )=P_{\text{new}}(x)P_{\text{new}}(\theta |x)$ with $%
P_{\text{new}}(x)=\delta (x-x^{\prime })$ fixed by the observed data while $%
P_{\text{new}}(\theta |x)=P_{\text{old}}(\theta |x)$ remains unchanged. We
see that in accordance with the minimal updating philosophy that drives the
ME method \emph{one only updates those aspects of one's beliefs for which
corrective new evidence (in his case, the data) has been supplied}\footnote{%
Use of a $\delta $ function has been criticized in that by implementing it,
the probability is completely constrained, thus it cannot be updated by
future information. This is certainly true! However, imposing one constraint
does not imply a revision of the other: An experiment, once performed and
its outcome observed, cannot be \emph{un-performed} and its result cannot be 
\emph{un-observed} by subsequent experiments.}\emph{.}

\section{Data and a moment}

\label{sec:3} In this general example, we extend our results from the
previous section. Again we wish to infer something about $\theta ,$ given
some information. The information that we are given in this example is some
observed data, $x^{\prime }$ and a constraint on the posterior in the form
of a moment. Here we apply the data constraint \emph{simultaneously} with
the moment constraint. Note that this problem cannot be solved by MaxEnt or
Bayes. For this example, we assume the constraints, 
\begin{equation}
\int dxd\theta P(x,\theta )=1~,  \label{Normalization}
\end{equation}%
which is our normalization constraint,%
\begin{equation}
\int d\theta P(x,\theta )=\delta (x-x%
%TCIMACRO{\U{b4}}%
%BeginExpansion
{\acute{}}%
%EndExpansion
)=P(x)~,  \label{Data}
\end{equation}%
which represents some observable data,%
\begin{equation}
\int dxd\theta P(x,\theta )f(\theta )=\left\langle f(\theta )\right\rangle
=F~,  \label{Expectation}
\end{equation}%
which represents some additional information. Maximizing the entropy given
the constraints with respect to $P(x,\theta )$ yields,%
\begin{equation}
P_{\text{new}}(x,\theta )=\frac{1}{Z}P_{\text{old}}(x,\theta )e^{\lambda
(x)+\beta f(\theta )}~,
\end{equation}%
where $Z$ is determined by using (\ref{Normalization}),%
\begin{equation}
Z=e^{-\alpha +1}=\int dxd\theta e^{\lambda (x)+\beta f(\theta )}P_{\text{old}%
}(x,\theta )
\end{equation}%
and the Lagrange multipliers $\lambda (x)$ are determined by using (\ref%
{Data})%
\begin{equation}
e^{\lambda (x)}=\frac{Z}{\int d\theta e^{\beta f(\theta )}P_{\text{old}%
}(x,\theta )}\delta (x-x%
%TCIMACRO{\U{b4}}%
%BeginExpansion
{\acute{}}%
%EndExpansion
)~.
\end{equation}%
The posterior now becomes%
\begin{equation}
P_{\text{new}}(x,\theta )=\frac{1}{\zeta (x,\beta )}P_{\text{old}}(x,\theta
)\delta (x-x%
%TCIMACRO{\U{b4}}%
%BeginExpansion
{\acute{}}%
%EndExpansion
)e^{\beta f(\theta )}~.  \label{Posterior-Both}
\end{equation}%
where $\zeta (x,\beta )=\int d\theta e^{\beta f(\theta )}P_{\text{old}%
}(x,\theta ).$

The Lagrange multiplier $\beta $ is determined by first substituting the
posterior into (\ref{Expectation})%
\begin{equation}
\int dxd\theta \left[ \frac{1}{\zeta (x,\beta )}P_{\text{old}}(x,\theta
)\delta (x-x%
%TCIMACRO{\U{b4}}%
%BeginExpansion
{\acute{}}%
%EndExpansion
)e^{\beta f(\theta )}\right] f(\theta )=F~,
\end{equation}%
which can be rewritten as%
\begin{equation}
\int dx\left[ \frac{1}{\zeta (x,\beta )}\int d\theta e^{\beta f(\theta )}P_{%
\text{old}}(x,\theta )f(\theta )\right] \delta (x-x%
%TCIMACRO{\U{b4}}%
%BeginExpansion
{\acute{}}%
%EndExpansion
)=F~.
\end{equation}%
Integrating over $x$ yields,%
\begin{equation}
\frac{\int d\theta e^{\beta f(\theta )}P_{\text{old}}(x^{\prime },\theta
)f(\theta )}{\zeta (x^{\prime },\beta )}=F
\end{equation}%
where $\zeta (x,\beta )\rightarrow \zeta (x^{\prime },\beta )=\int d\theta
e^{\beta f(\theta )}P_{\text{old}}(x^{\prime },\theta )$. Now $\beta $ can
be determined by%
\begin{equation}
\frac{\partial \ln \zeta (x^{\prime },\beta )}{\partial \beta }=F~.
\label{F}
\end{equation}

The final step is to marginalize the posterior, $P_{\text{new}}(x,\theta )$
to get our updated probability,%
\begin{equation}
P_{\text{new}}(\theta )=\frac{1}{\zeta (x^{\prime },\beta )}P_{\text{old}%
}(x^{\prime },\theta )e^{\beta f(\theta )}
\end{equation}%
Additionally, this result can be rewritten using the product rule as 
\begin{equation}
P_{\text{new}}(\theta )=\frac{1}{\zeta ^{\prime }(x^{\prime },\beta )}P_{%
\text{old}}(\theta )P_{\text{old}}(x^{\prime }|\theta )e^{\beta f(\theta )}~.
\end{equation}%
where $\zeta ^{\prime }(x^{\prime },\beta )=\int d\theta e^{\beta f(\theta
)}P_{\text{old}}(\theta )P_{\text{old}}(x^{\prime }|\theta ).$ The right
side resembles Bayes theorem, where the term $P_{\text{old}}(x^{\prime
}|\theta )$ is the standard Bayesian likelihood and $P_{\text{old}}(\theta )$
is the prior. The exponential term is a \emph{modification} to these two
terms. In an effort to put some names to these pieces we will call the
standard Bayesian likelihood the \emph{likelihood} and the exponential part
the \emph{likelihood modifier} so that the product of the two gives the 
\emph{modified likelihood}. The denominator is the normalization or \emph{%
marginal modified likelihood}.\footnote{%
Including an additional constraint in the form of $\int dxd\theta P(x,\theta
)g(x)=\left\langle g\right\rangle =G$ could only be used when it does not
contradict the data constraint (\ref{Data}). Therefore, it is redundant and
the constraint would simply get absorbed when solving for $\lambda (x)$.}

\section{The econometric problem}

\label{sec:4} This is a general example of an ill-posed problem using the above method: A
factory makes $k$ different kinds of bouncy balls. For reference, they
assign each different kind with a number, $f_{1},f_{2},...f_{k}$. They ship
large boxes of them out to stores. Unfortunately, there is no mechanism that
regulates how many of each ball goes into the boxes, therefore we do not
know the amount of each kind of ball in each or all of the boxes. However,
we are informed that the company does know the average amount of balls, $F$
in each of the boxes over the time that they have been in existence. What is
the probability of getting a particular kind of ball in one of the boxes? At
this point one could use MaxEnt to answer the question, assuming that the
'average' could be substituted for the moment constraint. Now let us
complicate the problem by suggesting that we would like a better idea of how
many balls are in each box (perhaps for quality control or perhaps the
customer would like more of one kind of ball than another). To do this we
randomly select a few balls, $n$ from a particular box and count how many of
each kind we get, $m_{1},m_{2}...m_{k}$ (or perhaps we simply open the box
and look at the balls on the surface). Now let us put the above example in a
more mathematical format.

Let the set of possible outcomes be represented by, $\kappa
=\{f_{1},f_{2},...f_{k}\}$ from a sample where the total number of balls, $%
N\rightarrow \infty $\footnote{%
It is not necessary for $N\rightarrow \infty $ for the ME method to work. We
simply wish to use the description of the problem that is common in
information-theoretic examples.} and whose sample average is $F.$ Further,
let us draw a \emph{data} sample of size $n,$ from the original sample whose
outcomes are counted and represented as $m=(m_{1},m_{2}...m_{k})$ where $%
n=\sum\nolimits_{i}^{k}m_{i}$. We would like to determine the probability
of getting \emph{any} particular outcome in one draw ($\theta _{i}$) given
the information. To discuss the probabilities related to this situation, we
implement observational data \emph{simultaneously} with an expectation
value. We start with the usual negative relative entropy for the joint space,%
\begin{equation}
S[P,P_{\text{old}}]=-\sum\limits_{m}\int d\theta ~P(m,\theta |n)\log \frac{%
P(m,\theta |n)}{P_{\text{old}}(m,\theta |n)}~.
\end{equation}%
We also have the following constraints,%
\begin{equation}
\sum\limits_{m}\int d\theta ~P(m,\theta |n)=1~,  \label{Ex4(normalization)}
\end{equation}%
\begin{equation}
P(m|n)=\int d\theta ~P(m,\theta |n)=\delta _{mm^{\prime }}~,
\label{Ex4(data)}
\end{equation}%
\begin{equation}
\sum\limits_{m}\int d\theta P(m,\theta |n)f(\theta )=\left\langle f(\theta
)\right\rangle =F~,  \label{Ex4(expectaion)}
\end{equation}%
where $\theta =(\theta _{1},\theta _{2}...\theta _{k}),$ $m=(m_{1}...m_{k})$
and $m^{\prime }$ is the observed data. Notice the use of the Kronecker for
the discrete case. Now we maximize the entropy given the constraints with
respect to $P(m,\theta |n)$ which yields,%
\begin{equation}
P_{\text{new}}(m,\theta |n)=P_{\text{old}}(\theta |n)\frac{P_{\text{old}%
}(m^{\prime }|\theta ,n)e^{\beta f(\theta )}}{\int d\theta e^{\beta f(\theta
)}P_{\text{old}}(\theta |n)P_{\text{old}}(m^{\prime }|\theta ,n)}~.
\end{equation}

We need to determine $P_{\text{old}}(m^{\prime }|\theta ,n)$ and $P_{\text{%
old}}(\theta |n)$ for our problem. The equation that we will use for the 
\emph{likelihood,} $P_{\text{old}}(m^{\prime }|\theta ,n)$ is simply the
multinomial distribution,%
\begin{equation}
P_{\text{old}}(m_{1}^{\prime }...m_{k}^{\prime }|\theta _{1}...\theta
_{k},n)=\frac{n!}{m_{1}^{\prime }!...m_{k}^{\prime }!}\theta
_{1}^{m_{1}^{\prime }}...\theta _{k}^{m_{k}^{\prime }}~.
\end{equation}%
Prior to receiving the information that the die is \emph{not} fair due to
the bias, we were completely ignorant of the status of the die. Therefore to
incorporate this ignorance we use a \emph{prior} that is flat, thus $P_{%
\text{old}}(\theta |n)=$ \emph{constant}. Being a constant, the prior can
come out of the integral and cancels with the same constant in the
numerator. (Also, the particular form of $P_{\text{old}}(\theta |n)$ is not
important for our current purpose so for the sake of definiteness we can
choose it flat for our example.)

Now we include our average information. To do this, we rewrite the moment
constraint (\ref{Ex4(expectaion)}) to reflect the special case by replacing
the function $f(\theta )$ with $\sum\nolimits_{i}^{k}f_{i}\theta _{i}$
where $f_{i}$ is a discrete parameter that reflects the label for the
outcomes and $F$ is the average. The sum relates the relationship of the
sides and $\theta _{i}$ is the continuous parameter that we wish to infer
something about. Thus the constraint is rewritten the following way.%
\begin{equation}
\sum\limits_{M}\int d\theta P(m_{1},\theta _{1}...m_{k}\theta _{k}|n)\left(
\sum_{i}^{k}~f_{i}\theta _{i}\right) ~\delta (\sum\limits_{i}^{k}\theta
_{i}-1)=F~,  \label{Ex4(expectaion2)}
\end{equation}%
where,%
\begin{equation}
\sum\limits_{M}=\sum\limits_{m_{1}=0}^{n}...\sum\limits_{m_{k}=0}^{n}%
\delta \left( \sum\limits_{_{1}}^{k}m_{i}-n\right) \quad \text{and}\quad
d\theta =d\theta _{1}...d\theta _{k}
\end{equation}%
Notice that $F$ reflects the \emph{average} relationship of the sides.

The resulting posterior is the product of the \emph{likelihood }and what we
have called the \emph{likelihood modifier}, $e^{\beta f(\theta )}$ or in
this case, $e^{\beta \sum\nolimits_{_{i}}^{k}f_{i}\theta _{i}}$ divided by
the normalization of the two,%
\begin{equation}
P_{\text{new}}(\theta _{1}...\theta _{k})=\frac{1}{\zeta }\delta
(\sum\limits_{i}^{k}\theta _{i}-1)\prod\limits_{i=1}^{k}e^{\beta
f_{i}\theta _{i}}\theta _{i}^{m_{i}^{\prime }}.  \label{Ex4(P)}
\end{equation}%
where $\zeta =\int d\theta \delta (\sum\nolimits_{i}^{k}\theta
_{i}-1)\prod\nolimits_{i=1}^{k}e^{\beta f_{i}\theta _{i}}\theta
_{i}^{m_{i}^{\prime }}$.

To determine $\beta $ we use (\ref{F}). This function can be complicated.
One may need to find a numerical solution for $\beta $ or an advanced search
technique such as Newton's method.

For simplicity we reduce the final $P_{\text{new}}(\theta )$ to $k-1$
dimensions,%
\begin{equation*}
P_{\text{new}}(\theta _{1}...\theta _{k-1})=\frac{1}{\zeta ^{\prime }}%
e^{\beta f_{k}\left( 1-\sum\limits_{i}^{k-1}\theta _{i}\right)
}(1-\sum\limits_{i}^{k-1}\theta
_{i})^{n-\sum\limits_{i}^{k-1}m_{i}}\prod\limits_{i=1}^{k-1}e^{\beta
f_{i}\theta _{i}}\theta _{i}^{m_{i}^{\prime }}~,
\end{equation*}%
where $\zeta ^{\prime }=\int d\theta e^{\beta f_{k}\left(
1-\sum\nolimits_{i}^{k-1}\theta _{i}\right)
}(1-\sum\nolimits_{i}^{k-1}\theta
_{i})^{n-\sum\nolimits_{i}^{k-1}m_{i}^{\prime
}}\prod\nolimits_{i=1}^{k-1}e^{\beta f_{i}\theta _{i}}\theta
_{i}^{m_{i}^{\prime }}$.

\subsection{Solving the normalization factor}

\label{sec:4.1} The denominator, $\zeta ^{\prime }$, which is the
normalization factor, can be a difficult integral. The general solution for
the $k$ sided die is a hypergeometric series which is calculated on a $k-1$
simplex,%
\begin{equation}
\zeta ^{\prime }=e^{\beta f_{k}}I_{1}(I_{2}(\ldots (I_{k-1})))\,,
\end{equation}%
where%
\begin{equation}
I_{j}=\Gamma (b_{j}-a_{j})\sum\limits_{q_{j}=0}^{\infty }\frac{\Gamma
(a_{j}+q_{j})}{\Gamma (b_{j}+q_{j})~q_{j}!}t_{j}^{q_{j}}I_{j+1}\quad \text{%
with}\quad I_{k}=1
\end{equation}%
and where $a_{j}=m_{k-j}^{^{\prime }}+1$, $b_{j}=n+j+1+\sum%
\nolimits_{i=0}^{j-1}q_{i}-\sum\nolimits_{i=0}^{k-j-1}m_{i}^{^{\prime }}($%
the terms $q_{0}$ and $m_{0}^{^{\prime }}=0)$, $t_{j}=\beta \left(
f_{k-j}-f_{k}\right) $, $\beta $ is the Lagrange multiplier and, $f_{i}$ and 
$f_{k}$ comes from $\Gamma (...)$ is the gamma function, and the terms $%
q_{0} $ and $m_{0}^{^{\prime }}=0.$ The index $j$ takes all discrete values
from $1 $ to $k-1$. The total number of counts or rolls of the die is $n,$
with $m_{i}^{^{\prime }}$ being the amount of counts for each parameter or
dimension, thus $n=\sum\nolimits_{i=1}^{k}m_{i}^{^{\prime }}.$ The summation
terms for each level of this nested series are represented by $q_{j}.$ The
factory information is codified in $t_{j}$, where $\beta $ is the Lagrange
multiplier and, $f_{i}$ and $f_{k}$ comes from (\ref{Ex4(expectaion2)}).

A few technical details are worth mentioning: First, one can have singular
points when $t_{j}=0$. In these cases the sum must be evaluated in the limit
as $t_{j}\rightarrow 0.$ Second, since $a_{j}$ and $b_{j}$ are positive
integers the Beta functions involve no singularities. Lastly, the sums
converge because $a_{j}>b_{j}$.

%%%%%%%%%%%%%%%%%%%%%%%%%%%%%%%%%%%%%%%%%%%%%%%%%%%%%%%%%%%%%%%%%%%%%%%%%%%%%%%%%%%%%%%%%%%%%%%%%%%%%%%%%%%%%%%%%%%%%%%%%%%%%%%%%%%%%%%%%	
\begin{figure}[!t]
\resizebox{1.0\columnwidth}{!}  {\includegraphics[draft=false]{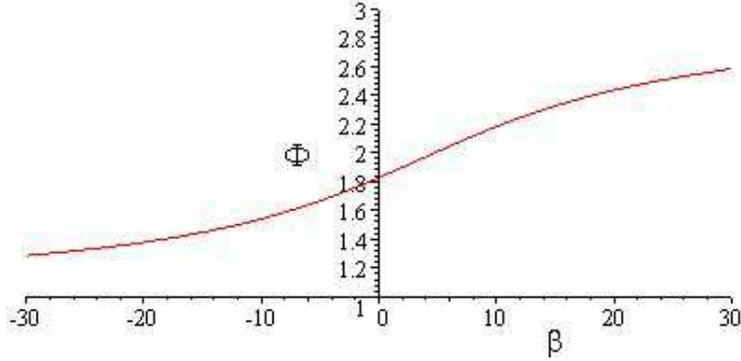}}  
\caption{This figure shows the relationship between $\protect\beta $ and $%
\Phi (\Phi =F(\protect\beta )).$ Notice that as the value for $\Phi $
approaches the extremities of the outcomes, $\protect\beta $ approaches
infinity.}
\label{Fig1:a}
\end{figure}
%%%%%%%%%%%%%%%%%%%%%%%%%%%%%%%%%%%%%%%%%%%%%%%%%%%%%%%%%%%%%%%%%%%%%%%%%%%%%%%%%%%%%%%%%%%%%%%%%%%%%%%%%%%%%%%%%%%%%%%%%%%%%%%%%%%%%%%%

\subsection{Numerical solutions}

\label{sec:4.2} We will extend the econometric example by applying the above
solutions to a specific problem where there are three kinds of balls labeled
1, 2 and 3. So for this problem we have $f_{1}=1,$ $f_{2}=2$ and $f_{3}=3.$
Further, we are given information regarding the average of all the boxes, $F.
$ For our example this average will be, $F=2.3.$ Notice that this implies
that on the average there are more $3$'s in each box. Next we take a sample
of one of the boxes where $m_{1}^{\prime }=11,$ $m_{2}^{\prime }=2$ and $%
m_{3}^{\prime }=7.$ The numerical solution for this example is,%
\begin{equation}
P_{\text{e}}(\theta _{1},\theta _{2})=\frac{1}{\zeta ^{\prime }}e^{\beta
(-2\theta _{1}-\theta _{2}+3)}\theta _{1}^{11}\theta _{2}^{2}(1-\theta
_{1}-\theta _{2})^{7}~,
\end{equation}%
where $\beta =14.1166$ and $\zeta ^{\prime }=1874.1247.$ We show the
relationship between $\beta $ and $F$ in Fig 1. The purpose of the Lagrange
multiplier is to enforce the moment constraint, therefore, as $F$ goes to
the extreme ($F\rightarrow 3$), $\beta \rightarrow \infty .$ This is
important to mention because it graphically illustrates that whether the
deviation from the sample mean is large or small, the ME method holds.

Another possible method suggested to use for this problem is the method of
types \cite{EMME}. This method essentially uses a form of Sanov's theorem,
which for this problem would be written as,%
\begin{equation}
P^{\ast }(\theta _{i})=\frac{Q(\theta _{i})e^{\eta _{i}f_{i}}}{%
\sum_{i}Q(\theta _{i})e^{\eta _{i}f_{i}}}~,
\end{equation}%
where $Q(\theta _{i})$ is "estimated" with the frequency of the data sample.
Thus $Q(\theta _{1})=\nu _{1}=11/20,$ etc. This produces the following
results:%
\begin{equation}
\theta _{t1}=0.3015,~\theta _{t2}=0.0971,~\theta _{t3}=0.6015_{.}
\end{equation}%
Taking the means of the ME solution yields,%
\begin{equation}
\left\langle \theta _{1}\right\rangle =0.2942,~\left\langle \theta
_{2}\right\rangle =0.1115,~\left\langle \theta _{3}\right\rangle =0.5942_{.}
\end{equation}%
Clearly the numerical solutions are very close, however, there are several
flaws with this large deviation method. The first is that $Q$ is treated as
a frequency. In the asymptotic case it would be appropriate to use a
frequency, unfortunately this is not that case. The data sample is finite, $%
n=20.$ Another flaw is that the method does not allow for fluctuations where
as the ME method does. Of course in the asymptotic case, fluctuations would
be ruled out, but again, this is not the case. There is an underlying theme
here: probabilities are not equivalent to frequencies \emph{except} in the
asymptotic case.

\section{Conclusions}

\label{sec:5} Using the ME method we were able to use information in the
form of data and moments to update our prior probabilities. A general
example was shown where the solution resembled the traditional form of Bayes
rule with the standard likelihood being modified by a factor resulting from
the moment constraint.

A specific econometric example was then solved in detail to illustrate the
application of the method. This case can be used as a template for real
world problems. Numerical results were obtained to illustrate explicitly how
the method compares to other methods that are currently employed. The ME\
method was shown to be superior in that it did not need to make asymptotic
assumptions to function and allows for fluctuations.

It must be emphasized that in the asymptotic limit, the ME form is analogous
to Sanov's theorem. However, this is only one special case. The ME method is
more robust in that it can also be used to solve traditional Bayesian
problems. In fact it was shown that if there is no moment constraint, one
recovers Bayes rule.

Therefore, we would like to emphasize that anything one can do with Bayes,
one can now do with ME. Additionally, in ME one now has the ability to apply
additional information that Bayesian methods could not. Further, any work
done with Bayesian techniques can be implemented into the ME method directly
through the joint prior. Finally the ME method can now also be used to solve
ill-posed problems in econometrics.

\bigskip

\noindent \textbf{Acknowledgements:} We would like to acknowledge valuable
discussions with A. Caticha, M. Grendar and C. Rodr\'{\i}guez.

\printindex
\end{document}